\documentclass[conference]{IEEEtran}
\IEEEoverridecommandlockouts
\usepackage{cite}
\usepackage{amsmath,amssymb,amsfonts}
\usepackage{xcolor}
\usepackage{booktabs}
\usepackage{threeparttable}
\usepackage{listings}
\usepackage{tikz}
\usetikzlibrary{positioning,arrows.meta,calc}
\usepackage{balance}
\usepackage{url}

\newcommand{\code}[1]{\texttt{\detokenize{#1}}}

\begin{document}

\title{SIMT-Aware Lockstep Verification and Functional-Coverage Closure
	Methodology for an Open-Source RISC-V GPGPU: A UVM 1.2 Environment}

\author{\IEEEauthorblockN{Samuel Moussa\IEEEauthorrefmark{1},
		Steven Ibrahim\IEEEauthorrefmark{1},
		Ahmad Sudky\IEEEauthorrefmark{1},\\
		Ahmad Fawzy\IEEEauthorrefmark{1},
		Abanoub Nabil\IEEEauthorrefmark{1},
		Alhassan Sayed\IEEEauthorrefmark{1},\\
		Hossam Hassan\IEEEauthorrefmark{2}, and
		Hyung-Min Yoon\IEEEauthorrefmark{3}}
	\IEEEauthorblockA{\IEEEauthorrefmark{1}Department of Electronics and
		Communications Engineering,\\
		Faculty of Engineering, Minia University, Egypt\\
		\IEEEauthorrefmark{2}Independent Researcher, South Korea \quad
		\IEEEauthorrefmark{3}Siliconarts, Inc., South Korea\\
		Corresponding author: samuel.moussa@mu.edu.eg}}

\maketitle

\begin{abstract}
	Open-source RISC-V GPGPUs such as Vortex ship with directed-kernel
	regressions but no industrial-grade verification: no reference-model
	checking, no functional-coverage model, and no sign-off discipline. This
	paper presents a complete UVM 1.2 environment and methodology that closes
	that gap, and reports what the resulting checking depth revealed about
	both the RTL and its reference model. The environment wraps a bus-master
	SIMT DUT with role-inverted agents, integrates Vortex's functional
	simulator (SimX) as a per-configuration golden model over DPI-C, and
	renders verdicts through two qualified checkers: a bidirectional
	end-state scoreboard and a per-instruction, per-lane lockstep comparator
	governed by five SIMT-specific alignment rules. A sound two-pass
	load-value feed makes racy, fenceless multi-core programs
	instruction-granularity verifiable (residual zero over 5{,}432
	retirements), with an explicitly characterized soundness boundary at
	asynchronous interrupt timing. A three-layer coverage model adds, to
	our knowledge, the
	first published SIMT functional-coverage layer for RTL GPU
	verification---warp divergence depth crossed with the
	immediate-post-dominator reconvergence stack, scratchpad bank conflicts,
	and memory-coalescing classes---closing the environment's own layers at
	98.1\% covergroup-bin /
	94.7\% total on the primary configuration (the third-party ISA layer
	closes separately at 83.1\% bins / 89.3\% weighted after
	register-index exclusions), using only
	machine-generated, RTL-cited exclusions, with a blocking waiver-integrity
	gate (the unintentional D-extension elaboration at this configuration
	adds no functional bins and only makes the code-coverage totals
	conservative; Section~\ref{sec:limits}). Every checker is proven able to fail by permanent fault-injection
	guards. The flow surfaced real defects on both sides of the comparison:
	an externally corroborated JALR LSB ISA deviation, a non-scaling watchdog
	constant (independently fixed upstream), an X-propagation window behind a
	reset relay found by restoring a silenced assertion, a missing AXI error
	path confirmed by fault injection, and a reference-model fetch bug found
	by the lockstep itself. We position the work against FuzzGPU (USENIX
	Security~'26), a concurrent RTL GPU fuzzer on the same DUT: the two are
	complementary---bug discovery by fuzzing versus coverage-quantified
	sign-off---and the JALR deviation was found independently by both. All
	findings ship in an evidence-cited register; every headline number in
	this paper carries its provenance, and the boundaries of the method are
	stated rather than waived.
\end{abstract}


\section{Introduction}
Open-source silicon has an industrial verification problem. The RISC-V
ecosystem has produced mature, publicly documented verification assets
for \emph{scalar} cores---step-and-compare lockstep environments,
constrained-random instruction generators, ISA coverage
libraries~\cite{corevverif, riscvdv, rvvi, imperasdv, difftest}---but the
most complex open processor class, the GPGPU, has none of this. Vortex,
the leading open RISC-V GPGPU~\cite{vortex-micro, vortex-carrv}, ships
with a directed-kernel regression and CI, no reference-model checking at
the RTL level, and no functional coverage model. Every consumer of
Vortex---research forks, GPGPU-PG studies, accelerator prototypes---inherits
that gap. Concurrently, FuzzGPU~\cite{fuzzgpu} demonstrated that
attackers and fuzzers can find real bugs in exactly this design space
(20 previously unknown bugs, 10 CVEs, across Vortex and Ventus), which
converts the verification gap from an academic inconvenience into an
exposure.

This paper presents a UVM 1.2~\cite{uvm} environment and methodology
that closes that gap for Vortex, and reports what the resulting
checking depth revealed about both the RTL and its reference model. Our
contributions:

\begin{enumerate}
	\item \textbf{A complete, configurable UVM environment for a SIMT
		      GPGPU} (Section~\ref{sec:env}): role-inverted agents (DUT = AXI master,
	      agents = reactive slaves), a DPI-C--integrated functional golden model
	      rebuilt per configuration, a passive RVVI-style observability layer,
	      and a bidirectional end-state scoreboard---any
	      $\textit{clusters}\times\textit{cores}\times\textit{warps}\times\textit{threads}$
	      topology from runtime options (plusargs) with elaboration-time parameter
	      self-checks.
	\item \textbf{Per-instruction lockstep for SIMT}
	      (Section~\ref{sec:lockstep}): five alignment rules---uuid-keyed
	      program-order sorting, mask-union record aggregation, per-lane compare
	      scoping, and volatile-CSR exclusion---that make CPU-style lockstep sound
	      for a SIMT machine, validated lane-exact at five configurations
	      (six programs)
	      spanning a $2\times2\times3\times2$ axis space.
	\item \textbf{A sound two-pass load-value feed}
	      (Section~\ref{sec:loadfeed}) that makes racy, fenceless multi-core
	      programs---architecturally undefined and previously unverifiable---%
	      instruction-granularity verifiable, with the \emph{residual as the
		      verdict} and non-vacuity proven by injection; plus a characterized
	      soundness boundary at asynchronous interrupt timing
	      (Section~\ref{sec:limits}).
	\item \textbf{A three-layer coverage model with honest closure}
	      (Section~\ref{sec:coverage}): a third-party ISA layer, an original
	      SIMT-microarchitecture layer---the first published functional-coverage
	      model of warp divergence depth, scratchpad bank conflicts, and
	      memory-coalescing classes for RTL GPU verification---and a
	      protocol/assertion layer; closed at 98.1\%/94.7\% (covergroup
	      bins/total) on the environment's own microarchitecture and
	      protocol layers---the third-party ISA layer closes separately
	      at 83.1\% bins / 89.3\% weighted after register-index
	      exclusions---using only machine-generated,
	      per-configuration, RTL-cited structural exclusions under a blocking
	      waiver-integrity gate, with unreachable-versus-unhit distinguished by
	      evidence and structural ceilings root-caused rather than waived.
	\item \textbf{An evidence-classified findings register}
	      (Sections~\ref{sec:rtlfindings}--\ref{sec:reffindings}): RTL bugs,
	      hazards, and observability limits, and reference-model defects---each
	      with \code{file:line} evidence and a disposition---including a
	      reference-model fetch bug found \emph{by} the lockstep, demonstrating
	      that a lockstep flow verifies the golden model as much as the DUT.
\end{enumerate}

Section~\ref{sec:positioning} positions the work against concurrent and
prior art; Section~\ref{sec:limits} delimits the method's own soundness
boundary; Section~\ref{sec:tapeout} separates what front-end functional
verification can claim from everything silicon sign-off would still
require.

\section{Background and Related Work}
\label{sec:positioning}

\subsection{Vortex}
Vortex~\cite{vortex-micro, vortex-carrv} is an open-source RISC-V GPGPU
organized as clusters $\to$ sockets $\to$ cores $\to$ warps $\to$
threads, with per-core instruction/data caches, optional shared L2/L3,
an immediate-post-dominator (IPDOM) reconvergence stack for divergence,
and hardware barriers. Execution units comprise ALU, FPU (FPnew), LSU,
SFU (CSR/warp control), and a tensor unit (TCU). The verified build is
RV32IMAF at RTL pin \code{7a52ee5}, primary configuration 1~cluster /
1~core / 4~warps / 4~threads (``1CL/1C/4W/4T'') with an AXI memory
interface, simulated on QuestaSim 2021.2~\cite{questa}. Notably, Vortex has \emph{no
	trap architecture}: there are no illegal-instruction, misaligned-address,
or CSR exceptions---a property with verification consequences developed
in Section~\ref{sec:rtlfindings}.

\subsection{Reference models}
Vortex ships SimX, a functional (non-timing) simulator maintained with
the RTL. SimX's emulator steps per instruction, returns a retirement
trace with full SIMT architectural state, and is decoupled from its
timing model---structurally an RVVI-shaped golden model~\cite{rvvi}.
Spike~\cite{spike} cannot execute the six SIMT instructions, so SimX is
the primary golden model and Spike serves as a \emph{secondary,
	independent} base-ISA cross-check. That audit has been carried out: on
the same linked ELF, Spike, SimX and the DUT retire exactly 11{,}076
architectural writebacks and agree on every program counter, destination
register and value---zero mismatches---with non-vacuity proven by
injecting a fault at one record and confirming it is named. Its scope is
bounded by construction and we do not overstate it: Spike is a scalar
ISS, so the audit covers warp~0 / lane~0 base-ISA execution only and
stops at the first Vortex custom operation. \textbf{The SIMT axis
	therefore has no independent reference}, and will not obtain one from
this direction. The consequence---that the primary golden model is
maintained by the same project and must itself be treated as a
verification target---is examined in Section~\ref{sec:reffindings}.

\subsection{Lockstep and reference-model flows}
Step-and-compare verification against a reference model is the standard
of care for RISC-V CPU cores: core-v-verif~\cite{corevverif} established
the open-source pattern with ImperasDV~\cite{imperasdv} over the
RISC-V Verification Interface~\cite{rvvi};
riscv-dv~\cite{riscvdv} supplies constrained-random programs and a
trace-compare co-simulation flow; DiffTest~\cite{difftest}
industrialized per-instruction architectural-state comparison over
DPI-C for the XiangShan processor line. Industrial-scale UVM
methodology on open silicon is demonstrated by
OpenTitan~\cite{opentitan}.
All of these target scalar (or
SIMD-vector) CPUs. Our environment descends from this methodology family
and extends it in three directions: to a bus-master DUT (role-inverted
agents), to SIMT semantics (per-lane masked compare, divergence,
multi-record retirements), and to weakly-ordered multi-core programs
(the two-pass load feed). Per-warp differential testing on this very DUT
has since also been demonstrated by FuzzGPU (below); the lockstep layer
presented here differs in its formalized alignment rules, its UVM
packaging, and its role as one of several qualified checkers inside a
sign-off flow rather than a fuzzer's oracle.

\subsection{GPU verification and fuzzing}
Software-level GPU kernel verification is a mature line:
GPUVerify~\cite{gpuverify} proves race- and divergence-freedom of
OpenCL/CUDA kernels; GKLEE~\cite{gklee} performs concolic testing and
explicitly targets divergent warps, non-coalesced accesses, and shared
memory bank conflicts as bug classes. These operate on kernel source,
not RTL; producer-consumer synchronization has been verified in GPU
programs~\cite{weft}, learned program generation has precedent in
compiler fuzzing~\cite{deepsmith}, and formal memory-consistency
verification reaches the RTL level in RTLCheck~\cite{rtlcheck} and the
PTX memory-model formalization~\cite{ptxmem}. At the hardware level, FuzzGPU~\cite{fuzzgpu} (USENIX
Security~2026) is the first RTL GPU fuzzer: it generates valid SIMT
programs (divergence-aware generation stack, deadlock-free barrier
planner, interleaved warp generation with taint propagation), runs them
on Verilator~\cite{verilator} models of Vortex and Ventus, and performs
trace-driven
differential testing against ISA-level instruction-set simulators with
per-warp state, reporting 20 previously unknown bugs and 10 CVEs. Its
evaluation is structural-coverage-based (line/toggle/expression), its
campaigns are time-boxed, and it carries no functional-coverage model,
no checker qualification, and no sign-off discipline---by design, since
it is a bug-discovery engine. Our work is the complementary half of the
problem: a UVM verification \emph{methodology} whose claims are
coverage-quantified, whose checkers are proven non-vacuous, and whose
exclusions are machine-generated and gate-checked. The two approaches
also cross-validate: the JALR LSB deviation we report as R1
(Section~\ref{sec:rtlfindings}) was independently found by FuzzGPU
(S2, upstream PR~\#339, CWE-682)---in the instruction-set-simulator
component of the same project; their RTL findings (X1--X3, PRs
\#356/\#358/\#359) are disjoint from ours, consistent with the
different pins and oracles involved. Concurrent with both, the Ventus
project developed GVM, a UVM-like
verification framework for its own OpenGPGPU design;
Vortex itself, to our knowledge, has had no public UVM environment prior
to this work.

\subsection{Checker qualification}
Mutation-based qualification of checkers---proving a checker can fail by
corrupting the design or its inputs---is established practice in CPU and
SoC verification, both academically~\cite{vacuity-date10} and
commercially; NVBitFI~\cite{nvbitfi} brought systematic fault injection
to GPUs for reliability studies. This work applies the discipline
end-to-end to a SIMT lockstep flow: every verdict-rendering checker
(Section~\ref{sec:verdicts}) is guarded by permanent fault-injection
regression tests, and the discipline is what allows the coverage and
pass-rate numbers of Section~\ref{sec:coverage} to be taken at face
value. Table~\ref{tab:positioning} summarizes the positioning of this
work against the reference-model verification and GPU testing flows.

\begin{table}[t]
	\caption{Positioning against reference-model verification and GPU
		testing flows. ``SIMT semantics'' covers per-lane masked compare and
		split-retire aggregation; ``checker qualification'' means
		fault-injection guards; ``waiver discipline'' means machine-generated,
		gate-checked exclusions. ISA coverage for our L1 layer is provided by
		the third-party riscvISACOV library; the SIMT microarchitecture model
		is this work's contribution.}
	\label{tab:positioning}
	\centering
	\scriptsize
	\setlength{\tabcolsep}{2.5pt}
	\begin{tabular}{lcccc}
		\toprule
		Property                   & \cite{corevverif,imperasdv} & \cite{difftest} &
		\cite{fuzzgpu}             & This work                                                                  \\
		\midrule
		Verification target        & scalar CPU                  & scalar CPU      & RTL GPGPU     & RTL GPGPU  \\
		Per-instruction compare    & \checkmark                  & \checkmark      & \checkmark    &
		\checkmark                                                                                              \\
		SIMT semantics             & ---                         & ---             & partial       & \checkmark \\
		Racy-program two-pass feed & ---                         & ---             & ---           & \checkmark \\
		ISA functional coverage    & \checkmark                  & ---             & ---           & \checkmark \\
		SIMT coverage model        & ---                         & ---             & ---           & \checkmark \\
		Checker qualification      & ---                         & ---             & ---           & \checkmark \\
		Waiver discipline          & ---                         & ---             & ---           & \checkmark \\
		Verdict taxonomy           & ---                         & ---             & ---           & \checkmark \\
		Framing                    & sign-off                    & agile DV        & bug discovery & sign-off   \\
		\bottomrule
	\end{tabular}
\end{table}

\section{Verification Environment}
\label{sec:env}

Fig.~\ref{fig:arch} shows the environment: active agents drive the
launch protocol and serve memory on the left, the functional reference
runs in lockstep on the right, and a passive probe layer feeds the two
checkers and coverage at the bottom.

\begin{figure*}[t]
	\centering
	\begin{tikzpicture}[
		font=\scriptsize,
		box/.style={draw, rounded corners=1pt, align=center, inner sep=3pt,
				minimum height=6mm},
		agent/.style={box, fill=gray!14},
		passive/.style={box, fill=gray!4, densely dashed},
		chk/.style={box, fill=gray!24},
		arr/.style={-{Stealth[length=2mm]}, semithick},
		tap/.style={-{Stealth[length=2mm]}, semithick, densely dashed}
		]
		\node[box, minimum width=46mm, minimum height=17mm] (dut)
		{\textbf{Vortex GPGPU (DUT)}\\
			clusters $\to$ sockets $\to$ cores\\
			warps $\times$ threads --- AXI \emph{master}};
		\node[agent, left=13mm of dut.west, anchor=east, yshift=8mm] (host)
		{host agent (active)};
		\node[agent, below=2.5mm of host] (dcr) {DCR agent (active)};
		\node[agent, below=2.5mm of dcr] (axi) {AXI agent\\(reactive slave)};
		\node[agent, below=2.5mm of axi] (mem) {memory model};
		\node[box, right=13mm of dut.east, anchor=west, minimum height=17mm] (simx)
		{\textbf{SimX} functional\\ reference (DPI-C)\\ per-config co-build};
		\node[passive, above=4mm of simx.north, anchor=south, minimum width=30mm] (spike)
		{\textbf{Spike} independent ISS\\ base-ISA audit (offline)\\ \emph{scalar: warp0/lane0}};
		\node[passive] (status) at ($(dut.south)+(-19mm,-10mm)$) {status agent};
		\node[passive] (cprobe) at ($(dut.south)+(0mm,-10mm)$)  {commit probe};
		\node[passive] (lprobe) at ($(dut.south)+(17mm,-10mm)$) {LSU probe};
		\node[passive] (rvvi) at ($(dut.south)+(9mm,-20mm)$)
		{RVVI monitor (analysis port)};
		\node[chk] (cov) at ($(dut.south)+(-27mm,-32mm)$)
		{functional\\ coverage};
		\node[chk] (lss) at ($(dut.south)+(3mm,-32mm)$)
		{lockstep scoreboard\\ per-instruction, per-lane};
		\node[chk] (ess) at ($(dut.south)+(38mm,-32mm)$)
		{end-state scoreboard\\ bidirectional, byte-exact};
		\draw[arr] (host.east) -- ++(4mm,0) |- ([yshift=5mm]dut.west);
		\draw[arr] (dcr.east)  -- ++(2.5mm,0) |- ([yshift=1mm]dut.west);
		\draw[arr, {Stealth[length=2mm]}-{Stealth[length=2mm]}]
		(axi.east) -- ++(2.5mm,0) |- ([yshift=-4mm]dut.west);
		\draw[arr] (axi) -- (mem);
		\draw[arr] ([yshift=4mm]dut.east) -- ([yshift=4mm]simx.west)
		node[midway, above, font=\tiny] {load/DCR/run};
		\draw[arr] ([yshift=-2mm]simx.west) -- ([yshift=-2mm]dut.east)
		node[midway, below, font=\tiny] {\shortstack{retirement\\records}};
		\draw[tap] ([xshift=-19mm]dut.south) -- (status.north);
		\draw[tap] (dut.south) -- (cprobe.north);
		\draw[tap] ([xshift=17mm]dut.south) -- (lprobe.north);
		\draw[arr] (cprobe.south) -- ([xshift=-9mm]rvvi.north);
		\draw[arr] (lprobe.south) -- ([xshift=8mm]rvvi.north);
		\draw[arr] (rvvi.south) -- ++(0,-2.5mm) -| (lss.north);
		\draw[arr] (rvvi.south) -- ++(0,-2.5mm) -| ([xshift=5mm]cov.north);
		\draw[arr] (status.south) -- ([xshift=-5mm]cov.north);
		\draw[arr] (simx.south) -- ++(0,-27mm) -| ([xshift=8mm]ess.north);
		\draw[arr] (simx.south) -- ++(0,-24.5mm) -| ([xshift=8mm]lss.north);
		\draw[arr] (mem.south) |- ($(ess.south)+(-4mm,-5mm)$)
		-- ($(ess.south)+(-4mm,0)$);
		\draw[arr, dotted] ([xshift=8mm]lss.south) -- ++(0,-8mm)
		-| ($(simx.east)+(5mm,0)$) -- (simx.east);
		\node[below] at ($(lss.south)+(28mm,-8mm)$)
		{load-value feed (pass 2, \S\ref{sec:loadfeed})};
		\draw[tap] (rvvi.east) -- ++(30mm,0) |- (spike.east)
		node[pos=0.25, right, font=\tiny, align=left]
			{\shortstack[l]{\code{+LOCKSTEP_TRACE}\\ (off by default)}};
	\end{tikzpicture}
	\caption{Environment architecture. Solid gray boxes are active UVM
		agents; dashed boxes are passive probes and offline tools (never
		checkers in the run); dark boxes are the two verdict-rendering
		scoreboards and coverage. The DUT is the bus master; stimulus is a
		compiled program in the memory model. \textbf{Two references, with
			different roles:} SimX is the \emph{primary} golden model, stepped live
		over DPI-C, and models SIMT; Spike is a \emph{secondary, independent}
		scalar ISS used offline on an exported retirement trace to cross-check
		the base-ISA subset only (warp~0 / lane~0, stopping at the first custom
		op). SimX is co-designed with the RTL and is therefore not independent;
		Spike is independent but cannot execute a SIMT kernel. Neither
		substitutes for the other, and the SIMT axis has no independent
		reference (\S\ref{sec:limits}).}
	\label{fig:arch}
\end{figure*}

\subsection{Role-inverted agent architecture}
A GPGPU DUT executes programs; it is not driven by them. The
environment therefore inverts the usual UVM roles. Five agents surround
the DUT: \emph{host} and \emph{device-configuration-register (DCR)}
agents (active) drive the launch protocol---kernel base address,
argument pointers, thread-count device-configuration registers---through
the platform's control interface; the \emph{AXI} agent is an active
\emph{responder}, a reactive slave backed by a memory model that
services the DUT's fetch and data traffic; a \emph{memory} agent covers
the non-AXI configuration of the same model; and a passive \emph{status}
agent observes busy/completion state, the retired-instruction count, and
the program counter. A virtual sequencer coordinates launch sequences.
Stimulus is therefore a \emph{program} (a compiled ELF), and the UVM
test class controls only \emph{how} it is launched, configured, and
perturbed---tests and programs are orthogonal and composed at the build
level.

\subsection{Golden-model integration}
SimX is linked into the simulation via DPI-C
(\code{simx_init}/\code{load}/\code{dcr_write}/\code{run} plus a
per-retirement record export). Both the RTL and SimX are rebuilt per
configuration so that DUT and reference always agree on topology.
Golden-model crashes are trapped by a signal handler and mapped to a
sentinel exit code, so a reference-model failure is \emph{classified}
(Section~\ref{sec:verdicts}) rather than crashing the run or silently
passing it.

\subsection{Passive observability layer}
All white-box visibility is bound, passive, and never a checker:
\begin{itemize}
	\item a \emph{commit probe} bound to the retire arbiter of every core,
	      capturing each retirement beat
	      $\{\textit{uuid}, \textit{wid}, \textit{PC}, \textit{rd},
		      \textit{wb}, \textit{tmask}, \textit{per-lane data}\}$---where
	      \textit{uuid} is the RTL's unique per-instruction identifier,
	      \textit{wid} the warp id, and \textit{tmask} the active-thread
	      mask---across all clusters, sockets, cores, and issue lanes;
	\item an \emph{LSU writeback probe} bound to the load-store slice,
	      capturing true per-lane load values after sign/zero extension (required
	      because load data is not observable at the commit arbiter,
	      Section~\ref{sec:rtlfindings});
	\item an RVVI-style interface and UVM monitor~\cite{rvvi} that publishes
	      the merged retirement stream through an analysis port;
	\item microarchitectural probes feeding the SIMT coverage layer:
	      scheduler, divergence/reconvergence, local-memory bank, memory
	      coalescer, cache-event, register-hazard, and commit-beat probes
	      (Section~\ref{sec:coverage}).
\end{itemize}
Lockstep capture is plusarg-gated; with the gate off, runs are proven
byte-identical to the plain environment.

\subsection{Checkers}
Two independent checkers render verdicts:
\begin{enumerate}
	\item \textbf{Bidirectional end-state scoreboard.} After completion,
	      every word the DUT wrote is compared byte-exact against SimX (forward
	      pass, up to 196{,}868 words in one cache-tier test), \emph{and} every
	      word SimX wrote that the DUT never wrote is checked (reverse pass,
	      catching dropped stores). Sub-word stores are handled with per-byte
	      validity masking; IEEE-rounding-legitimate FP differences are
	      tolerance-compared (and bounded, Section~\ref{sec:rtlfindings});
	      read-only/GOT sections are excluded by region.
	\item \textbf{Lockstep scoreboard} (Section~\ref{sec:lockstep}): every
	      retired instruction compared per active SIMT lane.
\end{enumerate}
A third, orthogonal gate makes the run verdict
\emph{assertion-aware}: any RTL runtime assertion firing fails the run
with a distinct exit code, and failing runs are excluded from coverage
merging. A dedicated negative test (a deliberate misaligned load whose
value is discarded) keeps this gate honest: it must fail through the
assertion path alone while the scoreboards pass.

\subsection{Configurability with self-checks}
One compiled environment runs any topology from plusargs
(\code{+CLUSTERS}/\code{+CORES}/\code{+WARPS}/\code{+THREADS});
interface widths (e.g., the 50-bit memory tag) are derived from the RTL
package, and elaboration-time assertions compare every UVM parameter
against its DUT counterpart, failing loud on mismatch. Validated
topologies include 1CL/1C/4W/4T, 1CL/2C, 2CL/2C/4W/4T, 4C/2W, and
8C/8W/2T.

\subsection{Protocol assertions}
An AXI4 SVA layer~\cite{sv} (${\sim}15$--$18$ protocol properties plus 11
handshake-stability assertions) checks burst legality, outstanding-count
consistency, and signal stability under backpressure inline on the DUT's
AXI interface; slave-side \emph{throttle} (ready wait-states) and
\emph{flood} (back-to-back R beats) stress modes---plusarg-gated and
proven byte-identical when off---exercise the stability properties. An
error-injection mode (\code{+AXI_INJECT_ERR}) drives \code{SLVERR}/%
\code{DECERR} responses on the read channel; its findings are reported
as R9/OBS-057 in Section~\ref{sec:rtlfindings}.

\section{Verdicts and Non-Vacuity}
\label{sec:verdicts}

Every run renders one of four verdicts: \textsc{pass},
\textsc{fail}, \textsc{unverifiable}, or (for lockstep runs)
\textsc{end-state-verified} with an instruction-granularity residual.
\textsc{unverifiable} is first-class: a run where the golden model
cannot render a verdict (e.g., a reference-model abort) is classified
with root-cause evidence---never force-compared, never silently
dropped, and never counted as a pass.

Verdicts are only as good as their ability to go red. The environment
therefore carries \emph{permanent fault-injection guards} as regression
tests, one per checker:
\begin{itemize}
	\item \textbf{wrong-value injection} (\code{+INJECT_FAULT}): one bit of
	      one DUT store is flipped; the end-state scoreboard must catch it;
	\item \textbf{dropped-store injection} (\code{+DROP_STORE}): one DUT
	      store is suppressed; the reverse scoreboard pass must catch it
	      (both guards historically detect at the exact address
	      \code{0x800075d8});
	\item \textbf{lockstep injection} (\code{+LOCKSTEP_INJECT}): one bit of
	      one retirement is flipped; the lockstep comparator must catch it at the
	      exact uuid/PC/lane;
	\item \textbf{DCR RAL injection} (\code{+DCR_RAL_INJECT}): a
	      register-abstraction-layer mirror corruption must be caught by the
	      register checker.
\end{itemize}
All guards must stay red on injection after any checker change---a
discipline directly in the tradition of mutation-based checker
qualification~\cite{vacuity-date10}, applied end-to-end here to a SIMT
lockstep flow. During the work, a nearly-accepted \emph{vacuous} gate
pass (negative tests reporting \textsc{pass} when their injection
plusargs were absent) was caught and codified into the rule: always
verify the injection message, never the verdict. This discipline, rather
than any single checker, is what allows the coverage and pass-rate
numbers below to be taken at face value.

\section{Per-Instruction Lockstep for SIMT}
\label{sec:lockstep}

\subsection{Architecture}
\begin{figure}[t]
	\centering
	\begin{tikzpicture}[
		font=\scriptsize,
		box/.style={draw, rounded corners=1pt, align=center, inner sep=2.5pt},
		eng/.style={draw, rounded corners=1pt, align=left, inner sep=3pt,
				fill=gray!8},
		arr/.style={-{Stealth[length=1.8mm]}, semithick}
		]
		\node[box, text width=35mm, fill=gray!12] (dut) at (0,0)
		{\textbf{DUT retirement stream}\\[1pt]
		\code{0x5c} mul \quad \code{0x59} lw \quad \code{0x5a} add\\[1pt]
		{\tiny retire order $\neq$ program order; one load splits into
		\code{0xd}/\code{0xe} records sharing a uuid; load data stale at the
		commit arbiter}};
		\node[box, text width=35mm, fill=gray!22, right=5mm of dut] (simx)
		{\textbf{SimX stream (DPI-C)}\\[1pt]
		\code{0x59} \quad \code{0x5a} \quad \code{0x5c}\\[1pt]
		{\tiny strict program order; uuid always 0; one record per
		instruction}};
		\node[eng, below=5mm of $(dut.south)!0.5!(simx.south)$, anchor=north,
			text width=78mm] (engine)
		{\textbf{Alignment engine --- five rules}\\[1pt]
			1~sort DUT records per (core, warp) by uuid\\
			2~key both streams by per-(core, warp) program order\\
			3~aggregate same-uuid records with thread-mask union\\
			4~loads: LSU-probe values + verifiable-region filter\\
			5~exclude machine-performance-counter CSR data};
		\node[box, below=3mm of engine.south, anchor=north, text width=78mm]
		(cmp) {\textbf{Per-lane compare}: PC $\cdot$ rd $\cdot$ value, active
			lanes only $\to$ matched / field mismatch / data mismatch / orphan,
			plus drain-empty on both sides};
		\draw[arr] (dut.south) -- ++(0,-1.5mm)
		-| ([xshift=-15mm]engine.north);
		\draw[arr] (simx.south) -- ++(0,-1.5mm)
		-| ([xshift=15mm]engine.north);
		\draw[arr] (engine.south) -- (cmp.north);
	\end{tikzpicture}
	\caption{The lockstep alignment pipeline. The DUT stream (left) carries
		the SIMT quirks that a scalar-CPU RVVI flow never meets; the reference
		stream (right) is strictly ordered. Each rule removes exactly one class
		of \emph{false} mismatch: sorting fixes out-of-order retirements
		(Rule~1), program-order keying bridges the reference's null uuids
		(Rule~2), mask-union aggregation reassembles split loads (Rule~3), the
		LSU probe with region filter restores sound load comparison (Rule~4),
		and volatile-CSR exclusion removes timing-only divergences (Rule~5).
		After alignment, any remaining difference outside the three declared
		exception classes is a real defect.}
	\label{fig:align}
\end{figure}
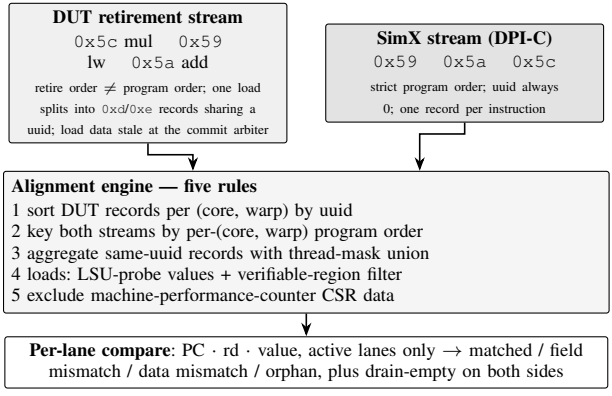

The DUT-side probe stream and the SimX per-retirement stream meet in a
lockstep scoreboard that aligns both streams per (core, warp) and
compares each retirement per active SIMT lane: PC, destination
register, and written value, with a four-way outcome taxonomy (matched /
field mismatch / data mismatch / orphan) and drain-empty checks on both
sides so that dropped or extra retirements are caught
(Fig.~\ref{fig:align}).

\subsection{Five SIMT-specific alignment rules}
\label{sec:rules}
A scalar-CPU RVVI flow never meets the following; each rule was forced
by simulation evidence and is, we believe, transferable to any SIMT
lockstep implementation.

\subsubsection{Retire order is not program order}
Within one warp, the commit arbiter retires whichever execution unit is
ready first (\code{VX_commit.sv:56-71}); an earlier-issued instruction
was observed retiring after two later ones. The functional model retires
in strict program order. \emph{Rule: align by the per-warp issue counter
	(uuid), sorted---never by position or cycle.}

\subsubsection{The golden model's uuid is not a cross-key}
SimX leaves its retirement uuid at zero (Section~\ref{sec:reffindings}),
so DUT and reference uuids cannot be matched directly. \emph{Rule: the
	alignment key is per-(core, warp) program order; the DUT uuid's high
	bits---which encode $(\textit{CORE\_ID} \ll \textit{NW\_BITS}) +
		\textit{wid}$ (\code{VX_uuid_gen.sv:40})---supply the (core, warp)
	attribution with no RTL change.}

\subsubsection{One instruction is not one retirement record}
A single load can appear as several commit records sharing one uuid
with partial, overlapping thread masks (observed: masks \code{0xd} then
\code{0xe}), because the LSU commits lanes as their memory responses
arrive---distinct from clean SIMD-beat splitting. \emph{Rule: aggregate
	DUT records by uuid with thread-mask union before comparing; never
	aggregate by start/end-of-packet flags.}

\subsubsection{Load data needs its own tap and its own soundness filter}
The commit arbiter's data field is stale for loads (the LSU writes the
register file on its asynchronous response path). A dedicated LSU
writeback probe recovers true per-lane values---but a naive compare is
\emph{unsound}: loads of uninitialized stack or local memory
legitimately differ between models (429 false mismatches on a known-good
kernel). \emph{Rule: compare a load lane only when the reference-model
	effective address lies in the verifiable data region and the golden
	value is not the initialization poison; defer everything else to the
	end-state check.} With this filter the load compare is exact and on by
default (74 in-region lanes compared, 113 filtered, 0 false mismatches
on the reference kernel).

\subsubsection{Performance-counter CSRs are model-divergent by definition}
\code{mcycle}/\code{minstret}/\code{mhpmcounter*} differ between a
timing-accurate DUT and a functional model (observed off-by-one).
\emph{Rule: the golden model flags the machine-performance-monitor CSR
	range as volatile and the comparator excludes their data (PC and
	destination still checked)---the standard RVVI exclusion class.}

Beyond the five rules, the comparator declares three bounded, logged
exception classes: the load-value deferral of Rule~4, the volatile-CSR
exclusion of Rule~5, and an \emph{fsqrt-only tolerance}. The DUT's
\code{fsqrt.s} is one ULP off correctly-rounded IEEE~754
(Section~\ref{sec:rtlfindings}), so an \code{fsqrt} writeback is
tolerated iff the two values are same-sign, finite, and within one ULP
(\code{fp32_within_ulp}, keyed on an \code{is_fsqrt} flag exported by
the reference); every toleranced comparison is tallied
(\code{n_fp_ulp_tol})---logged, never silent. All other floating-point
operations compare bit-exact, and an \code{fsqrt} beyond one ULP, or a
NaN/Inf mismatch, still fails. A two-pass reconvergence feed (armed
with the load feed on \code{+LOCKSTEP_LOADFEED}) then forces the
accepted DUT square-root result into the reference's FP register file
(re-NaN-boxed for the FLEN=32 build) and re-checks every downstream
operation bit-exactly, so the accepted deviation cannot cascade into
false errors; \code{fpu_test} and \code{fpu_mt} verify to residual
zero under this handling (1{,}675 matched retirements; four sqrt
reconvergences in the multi-threaded case). Each rule also carries a
DUT precondition an adopter must re-derive: Rule~1 a unique per-warp
issue counter that does not wrap within a run; Rule~2 a stable (core,
warp) attribution key; Rule~3 records that may split per lane but
share an identifier; Rule~4 a tap where true load values are
observable; Rule~5 knowledge of the model-divergent CSR ranges.

\subsection{Cross-configuration validation}
The lockstep was validated lane-exact---zero mismatches, zero
orphans---across a configuration matrix spanning
$\textit{NCL}\in\{1,2\}$, $\textit{NC}\in\{1,2\}$,
$\textit{NW}\in\{2,4,8\}$, $\textit{NT}\in\{2,4\}$
(Table~\ref{tab:matrix}), including a nested-divergence kernel
(asymmetric $3\text{v}1 \to 2\text{v}1 \to 1\text{v}1$ splits) that
exercised the mask-union aggregation through divergence and
reconvergence with no comparator changes.

\begin{table}[t]
	\caption{Lockstep configuration matrix (all lane-exact: 0 mismatches,
		0 orphans). Tallies are recorded in the committed verification
		documents; per-run raw logs were not retained.}
	\label{tab:matrix}
	\centering
	\small
	\begin{tabular}{lr}
		\toprule
		Configuration                    & Matched writebacks \\
		\midrule
		1CL/1C/4W/4T (vector-add kernel) & 1{,}035 / 1{,}035  \\
		1CL/1C/4W/4T (nested divergence) & 2{,}668 / 2{,}668  \\
		1CL/2C/4W/4T                     & 1{,}801 / 1{,}801  \\
		2CL/2C/4W/4T                     & 3{,}333 / 3{,}333  \\
		1CL/1C/2W/2T                     & 855 / 855          \\
		1CL/1C/8W/4T                     & 1{,}423 / 1{,}423  \\
		\bottomrule
	\end{tabular}
\end{table}

\section{Verifying Racy Programs: the Two-Pass Load-Value Feed}
\label{sec:loadfeed}

\begin{figure}[t]
	\centering
	\begin{tikzpicture}[
		font=\scriptsize,
		box/.style={draw, rounded corners=1pt, align=center, inner sep=3pt},
		arr/.style={-{Stealth[length=1.8mm]}, semithick}
		]
		\node[box, fill=gray!10, text width=78mm] (p1)
		{\textbf{Pass 1 --- live lockstep}\\[1pt]
			20 racy loads $\to$ 138 cascaded divergent retirements\\
			(5{,}314 matched, 118 data mismatches, of 5{,}432)};
		\node[box, text width=78mm, below=3.5mm of p1] (feed)
		{\textbf{Feed}: $(\textit{core}, \textit{warp}, \textit{PC},
			\textit{occurrence}) \mapsto$ DUT loaded value\\[1pt]
		{\tiny racy loads only; every other retirement stays
		hard-compared}};
		\node[box, fill=gray!10, text width=78mm, below=3.5mm of feed] (p2)
		{\textbf{Pass 2 --- reference replay follows the feed}\\[1pt]
		\textbf{residual 0} over 5{,}432/5{,}432 retirements};
		\node[box, text width=78mm, below=3.5mm of p2, fill=red!6] (bnd)
		{\textbf{Soundness boundary}: interrupt timing collapses
		116~$\to$~7, not 0\\[1pt]
		{\tiny keying-independent: the models take interrupts at different
		instruction boundaries}};
		\draw[arr] (p1) -- (feed);
		\draw[arr] (feed) -- (p2);
		\draw[arr] (p2) -- (bnd);
	\end{tikzpicture}
	\caption{The two-pass load-value feed on the flagship fenceless
		two-core case. Pass~1 runs the live lockstep and records the DUT's
		resolution of every racy load, keyed by (core, warp, PC, occurrence) so
		that replays with different instruction interleavings still address the
		same load site. Pass~2 re-executes the reference model while feeding it
		those values, so both models share each race outcome and the residual
		becomes the verdict: zero mismatches over all 5{,}432 retirements.
		Feeding is not suppression---a difference not explained by a fed load
		remains a hard error. Asynchronous interrupts sit outside the fixed
		point: the residual there is proven keying-independent and is reported
		as classified, not forced green.}
	\label{fig:feed}
\end{figure}
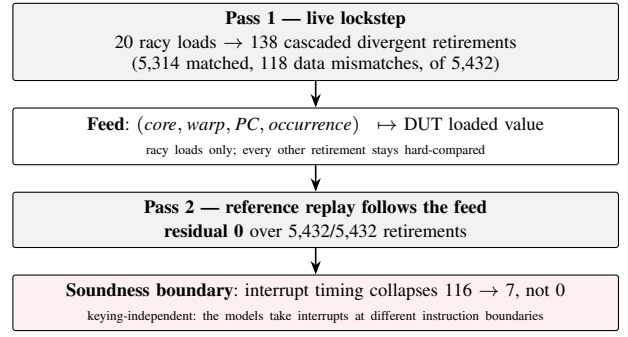

Fenceless multi-core programs have no architecturally defined result on
a weakly coherent machine, so a per-instruction comparison against a
functional model diverges legitimately: the two models resolve the same
race differently. Rather than waive such programs, we make them
verifiable (Fig.~\ref{fig:feed}). A first pass records the DUT's
per-lane loaded values at the
racy load sites, keyed by (core, warp, PC, occurrence); a second pass
re-runs the reference following those values, so the models share the
race outcome and any \emph{remaining} difference is a real defect. This
is a load-value feed in the RVVI sense, and it is not suppression: a
residual not explained by a fed load stays a hard error.

On the flagship case---20 racy loads cascading into 138 divergent
retirements at first pass---the residual collapses to \textbf{zero over
	5{,}432 retirements} (5{,}314 matched, 118 data mismatches at pass one;
5{,}432/5{,}432 matched at pass two), upgrading the verdict from
``\textsc{unverifiable}'' to \emph{verified modulo the fed race
	resolutions}. The feed's soundness rests on an assumption we state
rather than discharge in general: that a divergent in-region load is a
genuine race resolution, not a DUT load-data defect. For the flagship
case the assumption was confirmed by trace analysis---the
first-divergence load reads a word still holding its pristine ELF
initialization value (nothing had overwritten it before the DUT's
read) in a single-hart program running on four shared-memory cores, a
genuine cross-core ordering race. The
boundary is precise and we state it rather than hide it: for an
\emph{interrupt} test the same feed collapses 116 divergences to 7 but
not to 0, because the two models take the interrupt at different
instruction boundaries, so a same-PC occurrence count itself diverges.
Re-keying the feed proved this is not an alignment artifact---the
residual is identical. Interrupt-timing divergence therefore remains
instruction-granularity unverifiable, while the end-state comparison
still passes.

\section{Stimulus}
\label{sec:stimulus}

\textbf{Directed kernels} (${\sim}30$, all compiled with the Vortex
toolchain, all byte-exact against the reference): arithmetic and memory
baselines; all 13 FPU operation classes; tensor-unit WMMA including
multi-warp collective launches; nested-divergence towers; barriers under
partial thread masks; \code{wspawn}/\code{tmc} sweeps; cache
miss-status-holding-register (MSHR) and memory-pressure stress; a
232\,KB-text kernel; a 256\,KB high-entropy store stress; division
corner cases; CSR writes under SIMT masks; and VOTE/SHFL operations.

\textbf{Constrained-random}: a riscv-dv~\cite{riscvdv} pipeline
targeting rv32im, with GPU-specific post-processing (machine-mode CSR
stripping, \code{ecall}$\to$\code{ebreak}) and end-state verification;
Twelve riscv-dv profiles were brought up; two are excluded as
\emph{unimplementable} on this DUT---%
unaligned-load/store (no misaligned support,
Section~\ref{sec:rtlfindings}) and illegal-instruction (no trap
architecture)---and one was later dropped as a byte-identical
duplicate of another profile, leaving nine generator profiles (twelve
seeded instances run in the suite). Jump-stream generation is
sanitized to keep the
generator's deliberate JALR-LSB probing away from an RTL deviation it
would otherwise trip (finding R1); the deviation itself is reported, not
hidden.

\textbf{SIMT-aware generation (simtgen)}: an original random C-kernel
generator targeting the microarchitectural axes that scalar generators
cannot reach. Its knobs are derived from the DUT's own configuration
space (warp/thread counts, LSU coalescer window, local-memory bank
count, IPDOM stack depth), informed by the generation-axis taxonomy
published by FuzzGPU~\cite{fuzzgpu} but implemented against our own
checking stack; no FuzzGPU code is used (an evaluation of reuse,
rejected for engineering reasons, is documented in the artifact).
\code{simtgen} implements four axes: divergence (structured branch
trees with controlled split depths and asymmetric masks), memory
(stride/scatter/bank-hostile access patterns), barrier topology
(single-warp barriers, deadlock-safe by construction; \code{wspawn} is
deliberately never emitted---it is a runtime-only bootstrap primitive
with its own structural waiver), and VOTE/SHFL (all eight intrinsics
per program). The barrier and VOTE/SHFL axes closed no new
coverage---their targets (\code{cp_vote_shfl_op} 8/8,
\code{cross_sfu_threads}) were already fully covered by dedicated
directed kernels (\code{vote_shfl}, \code{bar_masks},
\code{sfu_masks}); their contribution is generator completeness and
seed-diversity robustness evidence, the same category of result as the
seed farm below. simtgen's measured effect is quantified in
Section~\ref{sec:coverage}; a ten-seed sweep across all nine riscv-dv
profiles (90 additional distinct programs, content-hash verified,
all passing) produced \emph{no measurable functional-coverage gain}
(every category bit-identical except toggle at $+0.06\%$)---a robustness
result we report as such, not as coverage progress: coverage saturates
inside the stimulus's reachable region, and the remaining work is
stimulus of a different \emph{kind}, not more samples of the same kind.

\textbf{Protocol stress}: the AXI throttle, flood, and error-injection
modes of Section~\ref{sec:env}, plus host/DCR launch-space sweeps.

\section{Coverage Methodology and Results}
\label{sec:coverage}

\subsection{A three-layer coverage model}
The functional model is organized in three layers with different owners
and deliberately different blind spots, and their databases are never
merged:
\begin{enumerate}
	\item \textbf{L1 --- ISA coverage} (third-party riscvISACOV layer): 80
	      active covergroups over RV32I/F/M/Zicsr/Zifencei. Sampling
	      \emph{every active lane as a hart} (4{,}581 samples) was proven to cover
	      exactly the same bin set as lane-0-only sampling (1{,}677 samples) on
	      our kernels---an empirical demonstration that ISA coverage is blind to
	      the SIMT axis, which is why L2 exists. A closed gap-hunt (frozen
	      2026-09-06) leaves 78/80 covergroups real and 429/516 target bins hit
	      (83.14\%; 89.28\% weighted) after register-index exclusions; two
	      covergroups are permanently empty by construction.
	\item \textbf{L2 --- SIMT microarchitecture} (ours; the layer this paper
	      contributes): probe-sampled covergroups for warp divergence crossed with
	      IPDOM stack depth (\code{divergence_cg}), reconvergence, barrier scope
	      and events, \code{tmc}/\code{wspawn} behavior, scratchpad bank conflicts
	      (\code{lmem_bank_cg}), memory-coalescing classes
	      (\code{coalesce_cg}), cache events, register hazards, and commit-beat
	      shapes (\code{beat_cg}), alongside instruction-class groups per
	      execution unit.
	\item \textbf{L3 --- system and protocol} (ours): SVA assertion and
	      directive coverage, AXI transaction space, DCR/host launch spaces, and
	      system-state crosses.
\end{enumerate}

\subsection{Closure discipline}
Three rules govern closure. (1)~\emph{Exclusions are structural,
	RTL-cited, and machine-generated}: a per-configuration generator emits
every waiver with a \code{file:line} citation (e.g., a write-response
stability assertion is unreachable because the adapter hardwires
\code{m_axi_bready = 1'b1}, \code{VX_axi_adapter.sv:313}), and a
blocking \emph{hits-invariant} merge gate verifies that every waived bin
had zero hits in the unwaived bank---on its first execution this gate
caught two real waiver defects, one of which had been active for weeks.
Exclusions are keyed to the configuration---e.g., the global-barrier
path is excluded only at single-core, where it is structurally
unreachable, and kept at $\geq 2$ cores. (2)~\emph{Reachable-but-unhit
	is left red}: bins that stimulus could reach but did not are reported
uncovered, never waived. (3)~\emph{Ceilings are root-caused}: the toggle
plateau was traced, per sub-module, to a read-only icache data port
(51{,}340 bins, 55.7\% sub-module toggle; 22{,}730 bins missing---26.4\% of
the \emph{entire} toggle gap---traceable to
\code{VX_socket.sv:106} tying \code{.WRITE_ENABLE(0)}) and constant
high address bits; an adversarial maximum-entropy kernel moved aggregate
toggle by $+0.02\%$, establishing the ceiling as structural. Coverage
banks are reported \emph{per configuration}: merging coverage databases
across topologies was shown to be invalid (instance-set inflation
deflates by-instance percentages).

\subsection{Results}
Table~\ref{tab:coverage} summarizes the banked configurations. All
totals are stated with their provenance and labels:

\begin{table}[t]
	\caption{Coverage banks (QuestaSim; per-configuration, never
		blended). All rows are the 2026-08-16 \emph{pre-relay-fix} banks;
		the reset-fixed 1CL bank (51 runs, 2026-08-18) totals the same
		94.7\% with branch 94.53\% and toggle 83.34\%---indicative only,
		two variables changed (\S\ref{sec:provenance}). A relay-fixed 2CL
		bank taken after this table was frozen (110 runs, 0 failures,
		93.4\% covergroup bins / 94.3\% total, on the grown
		post-remediation coverage model) restores cross-configuration
		comparability; the pre-fix columns are retained for provenance.}
	\label{tab:coverage}
	\centering
	\small
	\setlength{\tabcolsep}{4.5pt}
	\begin{threeparttable}
		\begin{tabular}{lrr}
			\toprule
			Metric                & 1CL/1C/4W/4T       & 2CL/2C/4W/4T      \\
			\midrule
			Covergroup bins (raw) & 370/377 = 98.1\%   & 989/1032 = 95.8\% \\
			Covergroup (weighted) & 99.8\%             & 99.5\%            \\
			Statement             & 98.1\%             & 98.3\%            \\
			Branch                & 95.1\%             & 95.7\%            \\
			Condition             & 90.4\%             & 88.8\%            \\
			Toggle                & 82.8\%             & 80.5\%            \\
			Assertion             & 96.9\%             & 98.9\%            \\
			Directive             & 100.0\%            & 100.0\%           \\
			\midrule
			\textbf{Total}        & \textbf{94.7\%}    & \textbf{94.6\%}   \\
			Runs passing          & 50/50$^{\ddagger}$ & 50/50             \\
			Coverage instances    & 2\,256             & 8\,275            \\
			\bottomrule
		\end{tabular}
		\begin{tablenotes}\footnotesize
			\item[$\ddagger$] \emph{Simulation runs}, not programs: 49
			distinct programs, one run twice under two bus modes.
			\item A third bank with shared L2/L3 enabled totals 93.2\%
			(51 runs, all passing), also on the pre-fix design.
		\end{tablenotes}
	\end{threeparttable}
\end{table}

\subsection{SIMT-layer closures and the generator's measured effect}
\begin{figure}[t]
	\centering
	\begin{tikzpicture}[font=\scriptsize]
		\draw[-{Stealth[length=1.6mm]}] (-0.1,0) -- (7.9,0);
		\draw[-{Stealth[length=1.6mm]}] (0,-0.1) -- (0,2.95);
		\node[rotate=90, font=\tiny, anchor=south] at (-0.55,1.4) {bins hit (\%)};
		\foreach \y/\l in {0/0, 0.6/25, 1.2/50, 1.8/75, 2.4/100}
			{\draw (-0.06,\y) -- (0.06,\y);
				\node[left] at (-0.08,\y) {\tiny \l};}
		\draw[dashed, gray] (0,2.4) -- (7.7,2.4);
		\fill[gray!50] (0.5,0) rectangle (1.1,1.8);
		\fill[black!75]  (1.2,0) rectangle (1.8,2.4);
		\node[above] at (0.8,1.8)  {\tiny 3/4};
		\node[above] at (1.5,2.4)  {\tiny 4/4};
		\fill[black!75]  (3.7,0) rectangle (4.3,2.4);
		\node[above] at (3.3,0.02) {\tiny 0/3};
		\node[above] at (4.0,2.4)  {\tiny 3/3};
		\node[align=center] at (1.15,-0.5) {\tiny \code{cp_split_depth}\\
			\tiny (IPDOM divergence)};
		\node[align=center] at (3.65,-0.5) {\tiny \code{cp_bank_conflict}\\
			\tiny (scratchpad banks)};
		\node[draw, fill=gray!50, inner sep=1.5pt, font=\tiny, anchor=west]
		at (3.0,2.8) {before};
		\node[draw, fill=black!75, inner sep=1.5pt, font=\tiny, anchor=west,
			text=white] at (3.9,2.8) {after simtgen};
	\end{tikzpicture}
	\caption{Effect of axis-directed generation on the two SIMT targets
		it genuinely closed (bins hit, as a percentage of the coverpoint's
		bin count). Before simtgen (light bars): depth-4 IPDOM divergence
		was unreachable by any of 50 random-program seeds (3/4 bins); bank
		conflicts never occurred in 70{,}498+ sampled cycles (0/3---itself
		root-caused to a coverpoint-definition defect, since the crossbar
		admits one winner per bank per cycle). Deliberate divergence-tower
		and bank-hostile kernels (dark bars) closed both to 100\%, now
		folded into a suite bank (2026-09-09). A third target previously
		reported here (coalescing classes) was re-examined and found
		already covered by the existing coalescer probe and kernels
		before simtgen existed---an attribution correction to our earlier
		isolated-merge reporting.}
	\label{fig:simtclosure}
\end{figure}
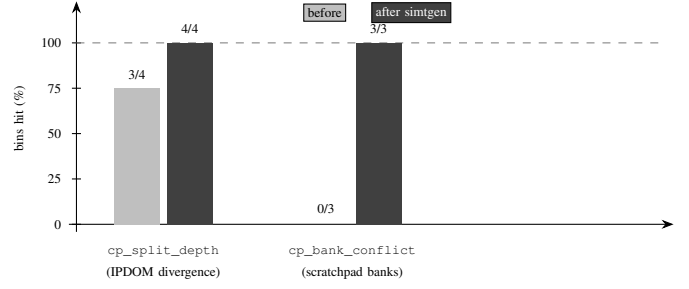

The SIMT layer is where the generator paid. Before simtgen, three
coverage targets were \emph{believed} open: the depth-4 IPDOM
divergence bin
(\code{divergence_cg.cp_split_depth}, 3/4---depth~3 never hit by any of
50 seeds), the bank-conflict bin of \code{lmem_bank_cg.cp_bank_conflict}
(0/3, never sampled in 70{,}498+ cycles), and the coalescing classes of
\code{coalesce_cg.cp_coalesce_kind}. Deliberate simtgen programs closed
the first two (Fig.~\ref{fig:simtclosure}): structured divergence
towers hit depth 4 (4/4), and bank-hostile access patterns (all lanes
of a warp addressing one bank by construction) surfaced real
conflicts; both closures are folded into a suite bank (2026-09-09).
Three honesty notes are attached to
this result and travel with it wherever it is cited:
\begin{itemize}
	\item the bank-conflict bin's initial ``gap'' was in fact a
	      \emph{coverage-model defect} (OBS-060): the coverpoint had been defined
	      over \emph{accepted} requests, but the local-memory crossbar admits
	      exactly one winner per bank per cycle---``two accepted requests on one
	      bank in one cycle'' was structurally impossible, and deliberate
	      bank-hostile stimulus proved it by producing zero conflicts. Reclassifying
	      the coverpoint onto \emph{offered} requests made the bin reachable and
	      simultaneously honest (169 conflict hits, 61 no-conflict, 70{,}460
	      idle); the closure credit is therefore shared between the probe fix
	      and the generator. A coverage definition must itself be validated
	      against the RTL's structure; the stimulus gap and the taxonomy defect
	      were disentangled by experiment.
	\item the coalescing classes, which an earlier isolated-merge report
	      attributed to simtgen, were re-examined and found already covered by
	      the existing coalescer probe and kernels before simtgen existed---an
	      attribution correction we report rather than quietly keep.
	\item the barrier and VOTE/SHFL generator axes closed nothing
	      because their targets were already fully covered by directed
	      kernels (Section~\ref{sec:stimulus}); their value is generator
	      completeness and seed-diversity robustness.
	      The marginal picture per generator kind: ninety additional
	      riscv-dv programs closed zero bins (the seed-farm result), while 57
	      simtgen programs (divergence and memory axes) closed the divergence
	      and bank-conflict targets---stimulus \emph{kind}, not volume, moves
	      this coverage model.
\end{itemize}

\subsection{Provenance of the design under verification}
\label{sec:provenance}
The design is an open-source register-transfer model pinned at a
specific revision, \emph{with eighteen locally modified files in the
	synthesizable register-transfer tree} (support and co-simulation glue
outside that tree is additionally modified, and the environment itself
is new code). Most changes are simulator-compatibility fixes made
during bring-up. We disclose this because a verification result is a
statement about a specific artifact, and ``upstream, unmodified'' would
not describe what we ran.

One of those modifications turned out to matter, and its history is
instructive. A library counter shipped with two assertions checking for
overflow and underflow. They fired during bring-up and were guarded
off---rewritten to skip whenever their inputs were unknown---so that the
environment could run at all. That guard remained for months. On
restoring the original assertion and root-causing why it fired, we found
a genuine defect (R10, Section~\ref{sec:r10}). We replaced the relay
with an asynchronous-assert, synchronous-deassert synchronizer; the
counter assertions then pass with no guard at all---twelve firings
become zero---and the local modification to that counter was retired in
favor of the unmodified upstream file.

Two consequences deserve emphasis. First, \textbf{the assertion had been
	correct}: what was silenced was a real property of the design, not a
testbench nuisance, and the cost was months of an unobserved
unknown-value window. Second, \textbf{fixing the defect reduced a
	coverage number}: branch coverage fell from 95.1\% to 94.5\% because
modules behind a relay had been executing their normal-operation paths
during the unknown-reset cycle, and those executions were \emph{counted
	as covered branches}. We report the lower figure as the correct one.
(Two variables changed between those measurements, so the delta is
indicative rather than isolated.)

\section{Findings I: RTL Observations}
\label{sec:rtlfindings}

Every finding is maintained in an evidence-cited register shipped with
the work (one entry per finding: what was seen, \code{file:line} and/or
run evidence, classification, disposition). Table~\ref{tab:rtlobs}
summarizes the RTL findings as R1--R10; two are developed below, and
further verified defects are noted afterwards.

\begin{table}[t]
	\caption{RTL findings (summary). The R1 spec deviation was
		independently found by FuzzGPU~\cite{fuzzgpu} in the ISS component
		(S2, PR~\#339); their RTL findings (X1--X3) are disjoint from
		ours; R2 was
		independently fixed upstream.}
	\label{tab:rtlobs}
	\centering
	\scriptsize
	\begin{tabular}{p{0.4cm}p{3.4cm}p{1.7cm}p{1.5cm}}
		\toprule
		ID  & Finding                                                                 & Class          & Disposition                                \\
		\midrule
		R1  & JALR target LSB not cleared; odd PC propagates via AUIPC                & Bug (ISA)      & needs RTL fix; stimulus sanitized          \\
		R2  & \code{STALL_TIMEOUT} uses \code{1**N} $\equiv$ 1; watchdog never scales & Bug (latent)   & fixed; \emph{independently fixed upstream} \\
		R3  & Misaligned access: no trap; silently retargeted/torn                    & Hazard         & expected per SW contract; gated            \\
		R4  & Core self-starts from reset; DCRs have no reset value                   & Hazard         & worked around (reset handshake)            \\
		R5  & Per-warp out-of-order commit                                            & Quirk          & handled (\S\ref{sec:rules})                \\
		R6  & One load $\to$ multiple commit records, overlapping masks               & Quirk          & handled (\S\ref{sec:rules})                \\
		R7  & uuid encodes flat core id + warp id                                     & Quirk          & exploited (\S\ref{sec:rules})              \\
		R8  & Load data not observable at commit arbiter                              & Observability  & closed (LSU probe)                         \\
		R9  & Write path fire-and-forget; \emph{no AXI error path}                    & Characteristic & cited; fault-injected (166)                \\
		R10 & Reset relay: reset registered in an unreset flop; one-cycle X window    & Bug (X source) & fixed; still upstream                      \\
		\bottomrule
	\end{tabular}
\end{table}

\subsection{An unknown reset for one cycle, found by restoring a
	silenced assertion (R10)}
\label{sec:r10}
\begin{figure}[t]
	\centering
	\begin{tikzpicture}[x=9mm, y=8.5mm, font=\scriptsize]
		\fill[red!12] (0,0.05) rectangle (1,5.5);
		\node[rotate=90, font=\tiny, align=center] at (0.5,2.8)
		{cycle 0: X window};
		\draw[semithick] (0,5) -- (0.75,5)
		\foreach \i in {1,...,6} { -- (\i,5.45) -- (\i+0.25,5) };
		\node[right, font=\tiny] at (7.35,5.2) {clk};
		\draw[semithick] (0,4.4) -- (7.3,4.4);
		\node[left, font=\tiny] at (-0.1,4.4) {reset};
		\fill[red!25] (0,3.05) rectangle (1,3.45);
		\node[font=\tiny\bfseries] at (0.5,3.25) {X};
		\draw[semithick] (1,3.4) -- (7.3,3.4);
		\node[left, font=\tiny] at (-0.1,3.25) {\code{reset_r}};
		\fill[red!25] (0,2.05) rectangle (1,2.45);
		\node[font=\tiny\bfseries] at (0.5,2.25) {X};
		\draw[semithick] (1,2.4) -- (7.3,2.4);
		\node[left, font=\tiny] at (-0.1,2.25) {\code{reset_o}};
		\draw[draw=red!60!black, fill=red!15, font=\tiny, align=center]
		(0,1.15) rectangle (1,1.65) node[pos=.5]
			{\emph{runs} (else branch)};
		\draw[fill=gray!12, font=\tiny] (1,1.15) rectangle (7.3,1.65)
		node[pos=.5] {held in reset};
		\node[left, font=\tiny] at (-0.1,1.4) {consumers};
		\draw[semithick, densely dotted] (0,0.4) -- (7.3,0.4);
		\node[left, font=\tiny] at (-0.1,0.4) {\code{reset_o}};
		\node[right, font=\tiny] at (7.35,0.4) {after fix};
	\end{tikzpicture}
	\caption{The reset-relay defect (R10) as a timing picture. The relay
		registers \code{reset} in a flip-flop that has no initial value and
		that nothing resets, so until the first clock edge (shaded cycle~0) its
		output---and therefore the reset of every module behind the
		relay---is unknown (X). Because \code{if (X)} takes the else branch,
		reset-conditioned consumer logic executes its \emph{normal-operation}
		paths during that window, and assertions evaluate on unknown operands:
		this is exactly what the (guarded-off) counter assertions had been
		reporting. With the async-assert/sync-deassert fix (bottom trace),
		\code{reset_o} asserts with reset from time zero and the twelve
		counter-assertion firings become zero---using the unmodified upstream
		file.}
	\label{fig:r10wave}
\end{figure}
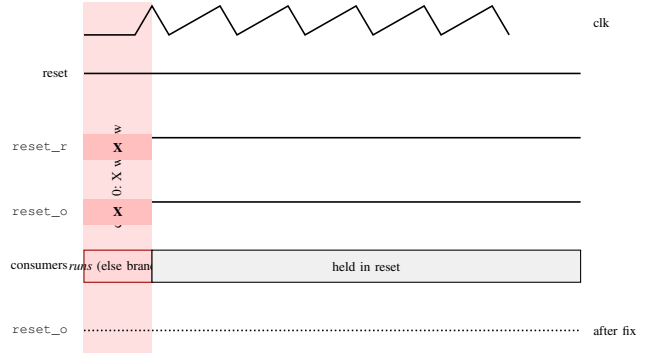

The design distributes reset through a relay module. That module
registers the incoming reset in a flip-flop which has no initial value
and which nothing resets:

\begin{lstlisting}
`PRESERVE_NET reg [R-1:0] reset_r;   // no initial value
always @(posedge clk) begin
    reset_r[i] <= reset;             // nothing resets THIS flop
end
assign reset_o[i] = reset_r[i / F];
\end{lstlisting}

Its output is therefore unknown from time zero until the first clock
edge (Fig.~\ref{fig:r10wave}), so every module instantiated behind a
relay observes an
\emph{unknown reset for one cycle}. A reset-conditional in such a module
takes its non-reset branch---\code{if (X)} is not true---so logic
intended to be held in reset executes, and any assertion inside that

branch evaluates on unknown operands.

The path to it is the point. A library counter in the same design ships
with assertions checking for overflow and underflow. Those assertions
fired during bring-up, at 5, 15 and 25\,ns, on the instruction- and
data-cache miss-status counters. They were guarded off---rewritten to
skip whenever their inputs were unknown---so that the environment could
run, and the guard stayed for months. Restoring the original assertion
and asking why it fired produced the diagnosis above. \textbf{The
	assertion had been correct the entire time}: it was reporting a real
property of the design, and what had been suppressed was the report, not
the problem.

We replaced the relay with an asynchronous-assert, synchronous-deassert
synchronizer, so that the relay output asserts whenever reset asserts
regardless of flip-flop state. The counter assertions then pass with no
guard at all---twelve firings become zero---and the local modification
to the counter was retired in favor of the unmodified upstream file. A
fifty-one-run regression passes with the assertions armed, and the
fault-injection guards were re-confirmed to fail on injection. The
defect is present in the current upstream release.

Two observations generalize beyond this design. First, \textbf{a
	checker disabled to make a bench run is a checker whose findings are
	lost}, and the loss is silent and open-ended: here it was months, and
the affected surface was every module behind a relay, not merely the one
that happened to carry an assertion. Second, \textbf{fixing the defect
	reduced a coverage number}
(Section~\ref{sec:provenance}): no coverage metric can report that some
of its own coverage came from a state that cannot legitimately arise.

\subsection{JALR does not clear the target LSB (R1)}
The RISC-V specification~\cite{riscv-spec} requires JALR to clear the
least-significant bit of the computed target. Vortex omits the clear
(\code{VX_alu_int.sv:222}: the branch destination is the raw
\code{rs1+imm}), and---having no trap architecture---cannot raise the
instruction-address-misaligned exception the specification prescribes
for a misaligned target on a non-compressed core. In the debug build
(\code{PC_BITS = XLEN}, identity PC conversion) the odd bit survives as
the \emph{architectural} PC. Fetch silently word-aligns
(\code{VX_fetch.sv:101}), so execution continues on correct instruction
words---but the skewed PC \emph{does} reach architectural results: every
\code{auipc}/\code{la} computes $rd = \textit{PC} + \textit{imm}$ and
inherits the skew, link-register writes accumulate it across chained
jumps (observed $+1 \to +3$), and downstream loads/stores go misaligned,
cascading into R3. The trigger is spec-legal stimulus: riscv-dv
deliberately sets the JALR base LSB expecting the architectural clear,
so roughly half of generated jumps derail---in a 12-profile suite, every
profile fired misaligned-access assertions (30--7{,}616 per run). In the
release build (\code{PC_BITS = XLEN$-$2}) a $+1$ target is masked away
by representation---spec-correct \emph{by accident}---while a $+2$
target would silently word-align where the specification demands a trap.
The fix is a one-line \code{\&\,\~{}1} at the destination adder; because
the reference model deliberately mirrors the no-clear behavior, the fix
must un-mirror both models together. FuzzGPU~\cite{fuzzgpu} found the
same deviation independently from a fuzzing flow entirely different
from ours (S2, upstream PR~\#339, CWE-682---filed against the project's
instruction-set simulator; their RTL findings X1--X3, at a different
pin, are disjoint from ours)---independent corroboration of the
\emph{deviation}, though not of its RTL embodiment.

\subsection{Further verified defects}
\textbf{fsqrt.s is one ULP off correctly-rounded IEEE~754~\cite{ieee754}}
(\code{OBS-014}): a lockstep lane-0 data mismatch (DUT
\code{0x3fef7750} vs.\ the SoftFloat-derived reference
\code{0x3fef7751}) localized by instruction and lane to the square-root
unit; the adjacent fdiv.s compared exact. The end-state scoreboard's FP
tolerance absorbs one ULP by design---\emph{only} the per-instruction
comparison surfaces it. \textbf{A second watchdog defect}
(\code{OBS-017}): beside R2's macro bug, the memory scheduler's
timeout localparam is denominated in simulation-time picoseconds,
yielding a ${\sim}1{,}000$-cycle budget on our timescale and flooding
L2/L3-enabled runs with 22{,}187 false-alarm messages while functionally
clean (2{,}789/2{,}789 lockstep match). \textbf{No AXI error path}
(\code{OBS-057}/R9): the AXI adapter hard-asserts that responses are
\code{OKAY}; injecting \code{SLVERR}/\code{DECERR} on every seventh
transaction produced exactly 166/166 assertion firings and no recovery
path---an evidence-based upgrade of a waiver into a measured design
limitation, consistent with the missing trap architecture.

\section{Findings II: Reference-Model Observations}
\label{sec:reffindings}

A central lesson of this work: \emph{a lockstep flow verifies the
	reference model as much as the DUT}. SimX is maintained by the same
project as the RTL, so shared assumptions can hide shared bugs; the
per-instruction compare surfaced defects in the model itself.

\subsection{Misaligned instruction fetch, found by the lockstep itself}
The reference read instructions at the exact byte program counter. A
computed indirect jump to an odd target---which the RTL tolerates,
because fetch word-aligns---made the reference read across an
instruction boundary and abort, classifying the run unverifiable. The
lockstep's own divergence report localized it. Word-aligning the fetch
while leaving the architectural PC odd (matching the DUT) recovered
those runs. The \emph{obvious} alternative---masking the JALR target in
the reference---was tried and \emph{reverted with evidence}: it produced
21{,}968 phantom PC mismatches, because it fixed the model to the
specification rather than to the DUT. This is the clearest case for
treating the golden model as a verification target in its own right: the
defect was in the reference, and only a per-instruction comparison could
have attributed it there.

\subsection{Abort surface and SIMT-divergent CSR semantics}
SimX's unknown-input behavior is to abort---56 \code{std::abort} sites
by disassembly (35 decode, 21 execute), the majority reachable through
the disassembly pretty-printer rather than the decoder core: a run can
be unverifiable because the golden model could not \emph{name} an
instruction. The lockstep upgrades such verdicts to ``byte-exact for all
17{,}664 retirements before the abort.'' Separately, a SIMT-divergent
CSR access produced 109 genuine data mismatches: the reference
serialized lane effects over the shared per-warp \code{fcsr} while the
RTL reads-broadcast and writes lane~0. The reference was corrected to
read-once/write-lowest-active-lane semantics; the RTL quirk is
deliberately \emph{not} mirrored, and is documented as a visible
architectural choice.

\section{Limitations and the Soundness Boundary}
\label{sec:limits}

\subsection{Asynchronous interrupt timing}
The interrupt-random test at two clusters collapses from 116 cascaded
divergences to 7 under the load feed---and does not reach zero. The
residual is proven \emph{keying-independent}: re-keying the feed from a
per-warp ordinal to $(\textit{cid},\textit{wid},\textit{PC},
	\textit{occurrence})$ leaves it identical, disproving feed-alignment
artifacts. The cause is genuine: the timing-accurate DUT and the
functional model take asynchronous interrupts at different instruction
boundaries, so an interrupt-affected path executes a different number of
times; no load-data feeding can align \emph{when} an interrupt fires.
The end-state compare against the post-feed model passes, so the
disposition is: end-state \textsc{verified}, instruction-granularity
residual classified---not forced green. This defines the method's
soundness boundary: \emph{two-pass trace replay is a fixed point for
	data-only divergence; asynchronous-input timing requires a
	step-follower reference}.

\subsection{Control-flow-steering races}
When racy loaded bytes steer control flow (a random jump test), pass-2
replay walks a different path and meets \emph{fresh} racy loads the
pass-1 trace never keyed (residual 15, thirteen of them load-class; the post-feed
end-state compare is clean). Two-pass replay has no fixed point when
races feed branches; an iterated (bounded fixed-point) feed or a
step-follower is required. The verdict is left honestly red.

\subsection{Structural blindness of differential testing}
Differential testing is structurally blind to a fault in the
\emph{stimulus}: DUT and reference execute the same binary, so a
mis-compiled kernel is mis-executed identically and every checker
passes while the kernel verifies nothing---observed concretely when a
tensor kernel's compile-time geometry silently mismatched the RTL
configuration (OBS-028/OBS-029). Equivalence validates only what
\emph{differs} between the models; everything shared (binary, memory
image, launch geometry) requires independent assertions. We closed
exactly one such input (kernel geometry asserted at elaboration,
injection-proven) and report the class as open.

\subsection{Other limitations}
\begin{itemize}
	\item \textbf{Reference-model independence.} SimX is not independent of
	      the DUT. The Spike audit covers the scalar subset only (warp~0 /
	      lane~0); \emph{the SIMT axis has no independent reference, and this is a
		      structural ceiling rather than an unfinished task.}
	\item \textbf{Simulator dependence.} All results are on one commercial
	      simulator (QuestaSim). Open-source simulation of UVM has matured
	      rapidly---recent Verilator elaborates upstream UVM~2017 with
	      constrained randomization---but SystemVerilog functional coverage
	      (covergroups) remains unsupported there, so the coverage-driven
	      half of this methodology cannot yet run on an open-source
	      simulator; the lockstep and end-state checking halves are, in
	      principle, portable today.
	\item \textbf{D-extension elaboration, scoped by inspection.} The D
	      extension is elaborated at the
	      primary configuration and MISA reports D=1 (\code{FLEN=64}, confirmed by
	      an elaboration probe) although the toolchain builds RV32IMAF and the
	      reference model models F only---unintentional-by-documentation as an
	      RTL fact. Its coverage impact is bounded by code inspection: the
	      coverage model's only D-specific bin (float--double conversion) is
	      config-aware-excluded at RV32 (\code{ignore\_bins} keyed on XLEN), the
	      ISA layer compiles no D plan, and no stimulus can emit a D
	      operation---so no functional bin is diluted. The
	      elaborated-but-unexecuted D logic sits only in code-coverage
	      denominators, which makes the reported totals \emph{conservative}
	      (they understate an F-only build), never inflated.
	\item \textbf{Per-mnemonic coverage soundness.} The frozen banks
	      predate the discovery that several instruction-class encodings alias
	      (\code{4'b1011} decodes as both a zero-conditional op and
	      \code{EBREAK}); per-mnemonic ALU identity in the frozen bank is
	      therefore unsound, and per-mnemonic claims are drawn only from
	      post-remediation measurements.
	\item \textbf{Structural coverage ceilings.} Toggle coverage plateaus
	      for root-caused structural reasons; we report the true value rather
	      than gaming it.
	\item \textbf{No trap architecture.} Exception-path verification is
	      unimplementable on this DUT; the corresponding generator profiles are
	      excluded as unimplementable rather than skipped silently.
	\item \textbf{Verified build.} Findings are stated against the debug
	      build at one RTL pin; the release build changes the visibility (not the
	      presence) of finding R1. The configuration-matrix tallies of
	      Table~\ref{tab:matrix} are documented results; their raw per-run logs
	      were not retained.
\end{itemize}

\section{Toward Tape-Out}
\label{sec:tapeout}

The work reported here is \emph{front-end functional verification} of an
RTL model. Silicon sign-off requires substantially more, and the two
categories below are deliberately separated because they demand different
tooling, different skills, and---critically---different claims. Neither
list is a statement of partial progress: every item is currently absent
unless noted.

\subsubsection{Remaining front-end (RTL functional) work}
(1)~Stimulus diversity, not volume---the seed-farm negative result
shows coverage saturating inside the scalar generator's reach; the
remaining work is stimulus of a different \emph{kind} (privileged
behavior, exception stimulus, bus errors---partly delivered by the
error-injection mode---and the two unimplemented simtgen axes).
(2)~Error and exception verification beyond the injection gate.
(3)~Formal property verification on arbitration-heavy control (cache
miss-status handling, commit arbitration), where dynamic simulation is
weakest and state space is small enough for proof; the IPDOM
reconvergence stack (push/pop balance, no underflow, reconvergence
restores entry mask) is the natural first target.
(4)~X-propagation and reset randomization---R10 proves the bug class
exists in this design, and several unreset elements have been
identified structurally but never exercised under randomized reset.
(5)~Configuration-matrix breadth beyond three sampled points.
(6)~Coverage-model provenance: the L2/L3 layers are self-authored
rather than traced to a specification document, so closure measures the
model, not the specification.
(7)~An independent SIMT reference---structurally absent, as above.

\subsubsection{Back-end and ASIC sign-off (entirely out of scope)}
Gate-level simulation with back-annotated timing; static timing
analysis across PVT corners; DFT (scan, ATPG, MBIST); CDC/RDC analysis
with metastability modeling; lint and structural sign-off; low-power
verification; physical-design closure; equivalence checking between RTL,
netlist, and post-layout; and a post-silicon path back into this
regression. A useful way to read the two lists: the front-end items
would raise the strength of the claims made in this paper, whereas the
back-end items are prerequisites for a different claim entirely---that
the design is manufacturable and will function in silicon. Nothing in
this work speaks to the latter.

\section{Conclusion}
\label{sec:conclusion}
We presented a UVM-based methodology that brings reference-model
verification---the standard of care for RISC-V CPU cores---to a SIMT
GPGPU, together with the first published SIMT functional-coverage model
for RTL GPU verification. The environment is complete and configurable:
role-inverted agents around a bus-master DUT, a per-configuration
co-built functional reference over DPI-C, a passive RVVI-style
observability layer, bidirectional end-state checking, and an
injection-proven non-vacuity discipline behind every verdict. The
lockstep layer contributes five SIMT alignment rules (each with a
stated DUT precondition) and a
two-pass load-value feed that turned an architecturally-undefined racy
multi-core program from unverifiable into verified modulo the fed race
resolutions (residual
0 over 5{,}432 retirements), while honestly delimiting its own soundness
boundary at asynchronous interrupt timing. The closure
methodology---machine-generated, gate-checked exclusions, root-caused
ceilings, residual bins dispositioned rather than waived---carried the
primary configuration to 98.1\%
covergroup bins / 94.7\% total, and the SIMT coverage layer was closed by an axis-directed
generator (the divergence and bank-conflict closures are folded into a
suite bank; the coalescing attribution and the coverage-model defect
found on the way are reported with their evidence).

The checking depth paid for itself in findings on both sides of the
comparison: in the RTL, a JALR ISA deviation with architectural
consequences (independently corroborated by concurrent fuzzing work), a
non-scaling watchdog macro (independently fixed upstream), a silent
misaligned-access corruption mode, a missing AXI error path, and a
reset-relay X window found by restoring a silenced assertion; in the
reference model, a misaligned-fetch bug that had been silently
discarding verification results, found by the lockstep itself. Both
registers ship with the work, each entry evidence-cited and
dispositioned. We believe the methodology---and its central lesson, that
the reference model must be verified alongside the DUT and the coverage
model alongside both---transfers directly to other open-source SIMT
designs, and fills the half of the open-GPU verification problem that
fuzzing does not address.

\section*{Acknowledgment}
The authors thank the Vortex research group at the Georgia Institute of
Technology for developing and openly maintaining the GPGPU platform on
which this work is built, and the maintainers of the open verification
assets (core-v-verif, riscv-dv, RVVI, riscvISACOV, Spike) leveraged by
the environment.

\balance

\end{document}